\documentclass{caosp310}

\usepackage{graphicx}

\usepackage{natbib}
\usepackage{amsmath}
\usepackage{comment}
\articleNo{X}
\pubyear{X}
\volume{X}
\volnumber{X}
\firstpage{1}
\received{Oct 1, 2025}
\accepted{XXX X, XXXX}

\def\BibTeX{{\rm B\kern-.05em{\sc i\kern-.025em b}\kern-.08em
             T\kern-.1667em\lower.7ex\hbox{E}\kern-.125emX}}

\begin{document}

%
\htitle{Oh thy mighty black holes!}
\hauthor{S. Panda {\it et al.}}

\title{Feeding frenzy in the mighty black holes:\\ what we could learn from them}


%
%
    \author{
        S. Panda\inst{1}\orcid{0000-0002-5854-7426}
      \and
        H. Benati Gon\c{c}alves\inst{2}\orcid{0009-0006-0492-9679}
      \and 
        T. Storchi-Bergmann\inst{2}\orcid{0000-0003-1772-0023}  
      \and
        M. \'Sniegowska\inst{3}\orcid{0000-0003-2656-6726}
      \and 
        B. Czerny\inst{4}\orcid{0000-0001-5848-4333}
      \and
        E. Bon\inst{5}\orcid{0000-0002-0465-8112}
      \and 
        P. Marziani\inst{6}\orcid{0000-0002-6058-4912}
      \and 
        N. Bon\inst{5}\orcid{0000-0002-3462-4888}
      \and 
        A. Rodr\'iguez Ardila\inst{7}\orcid{0000-0002-7608-6109}
      \and 
        D. May\inst{1}\orcid{0000-0002-2523-3551}
      \and
        M. A. Fonseca Far\'ia\inst{8}\orcid{0000-0002-7865-3971}
        \and
        L. Fraga\inst{8}\orcid{0000-0003-0680-1979}
      \and 
        F. Pozo Nu\~nez\inst{9}\orcid{0000-0002-6716-4179}
      \and 
        E. Ba\~nados\inst{10}\orcid{0000-0002-2931-7824}
      \and
        J. Heidt\inst{11}\orcid{0000-0002-0320-1292}
      \and
       K. Garnica\inst{12}\orcid{0000-0002-9596-9759}
      \and
       D. Dultzin\inst{12}\orcid{0000-0001-5756-8842}
       }

%
\institute{
           International Gemini Observatory/NSF NOIRLab, Casilla 603, La Serena, Chile, \email{swayamtrupta.panda@noirlab.edu}
         \and 
Departamento de Astronomia, Instituto de Física, Universidade Federal do Rio Grande do Sul, CP 15051, 91501-970, Porto Alegre, RS, Brazil\\
         \and
School of Physics and Astronomy, Tel Aviv University, Tel Aviv 69978, Israel\\
         \and
Center for Theoretical Physics, Polish Academy of Sciences, Al. Lotnik\'ow 32/46, 02-668 Warsaw, Poland\\
         \and
Astronomical Observatory Belgrade, Volgina 7, 11060 Belgrade, Serbia\\
         \and
INAF-Astronomical Observatory of Padova, Vicolo dell'Osservatorio, 5, 35122 Padova PD, Italy\\
         \and
Observat\'orio Nacional, Rua Jos\'e Cristino, 77, S\~ao Cristov\~ao, 20921-400, Rio de Janeiro, RJ, Brasil\\
         \and
Laborat\'orio Nacional de Astrof\'isica (LNA), Rua dos Estados Unidos 154, Bairro das Na\c c\~oes, Minas Gerais, Brazil\\
         \and
Astroinformatics, Heidelberg Institute for Theoretical Studies, Schloss-Wolfsbrunnenweg 35, 69118 Heidelberg, Germany\\
         \and
Max-Planck Institut f\"ur Astronomie, K{\"o}nigstuhl 17 Heidelberg, Germany\\
         \and
Landessternwarte, Zentrum f\"ur Astronomie der Universit\"at Heidelberg, K{\"o}nigstuhl 12, 69117 Heidelberg, Germany\\
         \and
Universidad Nacional Aut\'onoma de M\'exico, Instituto de Astronom\'ia, AP 70-264, 04510, CDMX, Mexico
          }
\date{Oct 1, 2025}

\maketitle

\begin{abstract}
Eddington ratio ($\lambda_{\text{Edd}}$) is a paramount parameter governing the accretion history and life cycles of Active Galactic Nuclei (AGNs). This short review presents a multi-faceted view of the importance of the Eddington ratio spanning varied AGN studies. We find that $\lambda_{\text{Edd}}$ is crucial for standardizing the Radius-Luminosity (R-L) relation — a necessary step for employing quasars (QSOs) as standardizable cosmological probes to help clarify the standing of the Hubble tension. In this data-driven era, we consolidated disparate aspects by developing novel relations borne out of large datasets, such as the robust, nearly \textit{universal anti-correlation} between fractional variability (F$_{\rm var}$) and $\lambda{_{\rm Edd}}$ derived from Zwicky Transient Facility (ZTF) data, which is vital for interpreting forthcoming high-cadence surveys like Rubin Observatory's LSST. Addressing the conundrum where JWST results suggest an overabundance of massive high-redshift black holes, we demonstrate that local AGNs offer clarification: Changing-Look AGNs (CLAGNs), driven by rapid $\lambda_{\text{Edd}}$ shifts, cluster in the low-accretion regime ($\lambda_{\text{Edd}} \sim 0.01$), a rate independently confirmed by our integral field spectroscopy and photoionization modeling of a well-known Seyfert 2 galaxy, rich in high-ionization, forbidden, coronal lines. Conversely, for the high-redshift, high-luminosity population where traditional reverberation mapping (RM) is highly impractical, photometric reverberation mapping (PRM) offers a rapid alternative to constrain accretion disk sizes, enabling efficient estimates of black hole masses ($\text{M}{_\text{BH}}$) and $\lambda{_\text{Edd}}$. Finally, we developed tailored semi-empirical spectral energy distributions (SEDs) for extremely high-accretion quasars, successfully validating their characteristic extreme physical conditions.\\

\end{abstract}

%
\section{Accretion rate - a key parameter in AGN studies}
\label{sec0}

The Black Hole Accretion Rate (BHAR) is among the fundamental properties that help define the state and activity of the central supermassive black hole (SMBH) residing at the very centers of galaxies. The BHAR or an equivalent parameter, the Eddington ratio, is the ratio of the net bolometric output from the active galactic nuclei (AGN) relative to the Eddington limit\footnote{L$_{\rm Edd} \approx 1.26 \times 10^{38} \left(\frac{{\rm M_{BH}}}{{\rm M_{\odot}}}\right)$ erg s$^{-1}$.}. Coupled with the knowledge of the mass of the SMBH, the Eddington ratio enables an understanding of the accretion history and life cycles of AGNs \citep{Netzer_2015, Padovani_2017, Alexander_2025, 2025Univ...11...69M}. The applications of the Eddington ratio are far-fetched, from studying the temporal modulation of radiative output from individual AGNs to constraining the bulk-statistical growth of AGNs across cosmic time. It has been found to be a direct correlate to explain the dispersion in the well-known broad-line region (BLR) radius - AGN luminosity (R-L) relation \citep{MLMA_2019, Du_Wang_2019, PandaMarziani2023FrASS}. The R-L relation is an empirical scaling relation, where the BLR radius is estimated using the reverberation mapping (RM) technique, where one can estimate the excess in the light travel time between photons that directly arrive from the source of ionizing radiation (in the case of the SMBH, this is the accretion disk - a flattened disk-like structure arising from the loss of angular momentum as the matter accretes onto the SMBH), and those that get intervened by gaseous media in-between accretion disk and the distant observer before eventually making their way to the observer. This offset in the light travel time between the two (photon) paths helps determine how far the intervening, line-emitting media is from the source of the radiation. This intervening medium, in our case, is the broad-line region (BLR), which produces the bulk of the emission lines that we observe in an AGN spectrum due to a variety of radiative processes, e.g., recombination, collisional excitation, and fluorescence \citep{Osterbrock_Ferland_2006}. We note, however, that the reverberation mapping technique has also been used to determine the sizes of the accretion disk itself (using the continuum reverberation mapping, \citealt{2015ApJ...806..129E}), the sizes of the dusty regions (dust reverberation, \citealt{2006ApJ...639...46S}), and to map the further-out narrow-line region (NLR, \citealt{2013ApJ...779..109P}).\\ 

Coming back to the R-L relation, having known the BLR radius, we now require the AGN luminosity. The general approach involves extracting the continuum flux from the AGN spectrum, wherein we look for narrow, line-free, continuum windows, and with an assumed cosmology, which gives us the luminosity distance, we can estimate the AGN luminosity. Therefore, having the AGN luminosity, for example, from a single-epoch spectrum, and with the aid of the R-L relation, we can infer the black hole masses using well-known scaling relations based on the virial theorem\footnote{here, the virial theorem requires the knowledge of the velocity dispersion (or FWHM) of a prominent emission line, its location (R$_{\rm BLR}$) from the central gravitational potential, i.e., the SMBH, and the assumption of the geometry and distribution of the emitting gas around the SMBH. This latter term is usually referred to as the virial factor or f-factor \citep{Collin_Kawaguchi_2006, Panda2019_Orientation}.} and extends to the creation of large quasar catalogs \citep{2003ApJS..145..199M, Vestergaard_Peterson_2006, Shen_2011, Panda2024_MCQ}. This, then, allows us to get insight into the black hole mass distribution over a range of redshift and infer their activity with the combined knowledge of the black hole mass and the net bolometric output. Another advantage of the R-L relation has been to utilize it as, cosmologically speaking, a standardizable relation to infer the cosmology using quasars \citep{Watson_2011, Haas_2011, Czerny_2013}. Here, we get the continuum flux as before, either through single epoch spectroscopy or carefully optimized photometry - either by a systematic modeling of the contaminants all except the accretion disk \citep{FPN2023MNRAS, Jaiswal_2025}, or through the usage of medium/narrow-band filters to avoid regions where such contributions from contaminants dominate \citep{Chelouche_2019, PandaFrancisco2024ApJL, FPN2025A&AL}. Instead of assuming a cosmological model, we then perform the reverberation mapping to extract the size of the emitting region, here R$_{\rm BLR}$, and infer the AGN luminosity using the R-L relation \citep{Blandford_1982, Peterson_2004, Bentz_2013, Panda2019FrASS, Cackett_2021}. Having uniquely determined the continuum flux and the luminosity from independent methods, we can then derive the luminosity distance for each of these objects\footnote{Given, R$_{\rm BLR} \propto$ L$^{0.5}$ from photoionization theory and the empirical R-L relation, and the AGN flux, F = L/ 4$\pi$d$_{\rm L}^2$, we can re-write luminosity distance, d$_{\rm L}$ $\propto$ R$_{\rm BLR}$/$\sqrt{4 \pi {\rm F}}$.}, and populate the Hubble-Lemaitre diagram and constrain the existing cosmological models. With the aid of quasars, we are well-poised to bridge the gap between the local and early Universe cosmological probes and help clarify the stance on the Hubble tension \citep{Czerny_2023, PandaMarziani2023FrASS, DiValentino_2025}. However, to infer the Hubble constant, we aggregate the RM AGNs with other cosmological probes, e.g., Type-1 Supernovae, chronometric measurements of the Universe expansion, baryon acoustic oscillation data, quasar angular sizes, H {\sc ii} starburst galaxies, and Amati-correlated gamma-ray burst data \citep{2022MNRAS.516.1721C, Czerny_2023}. On the other hand, to get to the Hubble constant directly from AGNs alone, we need the aid of standard candles or standard rulers, e.g., the sizes of the accretion disks allow us to characterize the angular sizes of the AGNs, in conjunction with the redshift-independent luminosity distances \citep[see e.g.,][]{Jaiswal_2025}.\\

A fundamental issue using the R-L relation has been the realization of the highly-accreting AGNs and their location on the R-L relation \citep{Du_2016, Grier_2017, Du_2018, MLMA_2019}. When the R-L relation was first established, the scaling relation was built using multi-epoch, ground-based, and space-based spectrophotometric observations of well-known nearby Seyfert galaxies \citep{Kaspi_2000, Bentz_2009}. The slope of the relation was found to be close to the expectation from the standard photoionization theory, i.e., $\sim$0.5, especially after careful host subtraction \citep{Wandel_1999, 2013ApJ...771...31N, 2022FrASS...950409P}. Fast forward to mid-2015, newer reverberation mapping (RM) campaigns, with luminous sources accreting at relatively higher rates albeit with lower variability amplitudes, started to populate a region in the R-L parameter space which demonstrated a significant dispersion from the existing R-L relation, thus giving rise to studies to understand what made these sources inherently different from their local counterparts \citep{Du_2018, MLMA_2019, Panda2019FrASS}. Later, in \citet{MLMA_2019}, we found the key reason for the dispersion was directly connected to the Eddington ratio - the higher the value, the more the source deviates from the scaling relation, especially in the direction where the inferred R$_{\rm BLR}$ is shorter than expected. The problem, however, with the Eddington ratio in this context, is the way it is derived - the bolometric luminosity is derived from the continuum luminosity, and so is the black hole mass. This is the same luminosity that is the `L' in the R-L relation, thus creating a circular loop. We need an independent observable that could trace the Eddington ratio. Many studies have pointed to the strength of the Fe {\sc ii} emission (or, R$_{\rm Fe}$) to be a viable surrogate to the Eddington ratio through carefully statistical and theoretically-motivated studies \citep{Sulentic_2000, 2003MNRAS.345.1133M, Marziani_2014, Shen_Ho_2014, Marziani_2018, Panda2019_Orientation, MLMA_2021}; c.f. \citet{Panda2024FrASS} for a recent overview. \citet{Du_Wang_2019} (see also, \citealt{PandaMarziani2023FrASS}) eventually made the breakthrough, and provided us with an R$_{\rm Fe}$-corrected R-L relation which not only solved the circularity problem, but brought the scatter in the newfound R-L to a measly 0.19 dex, re-instating the R-L relation as a standardizable relation for use in cosmology. \\

This is just one of the many highlights of the Eddington ratio and how it has come to the rescue. In later sections of this short contribution, we demonstrate a few recent studies that involve this key parameter and how it has helped improve our understanding of AGNs, and set the stage for the ongoing and upcoming, exciting studies with state-of-the-art facilities and observatories, e.g., JWST \citep{JWST_2006SSRv..123..485G}, Dark Energy Spectroscopic Instrument \citep[DESI,][]{DESI_2025arXiv250314745D}, and Rubin-LSST \citep{LSST2019ApJ, FPN2023MNRAS, Czerny2023A&A}. In the following sections, we will summarize a few recent advancements that have allowed us to unravel the connection between the black hole activity and the mass of the SMBH, and in turn, the Eddington ratio, as well as the novel relations we have found. That would not be possible without the use of spectrophotometric observations of AGNs using a multitude of telescope facilities - targeted and survey-mode, single-epoch and monitoring campaigns, archival and new observations, covering the two hemispheres.  

\section{Getting ready for LSST: Photo-variability reveals a nearly Universal scaling relation}
\label{sec1}

We have all been greeted with the \textit{first look} of Rubin\footnote{\url{https://rubinobservatory.org/gallery/collections/first-look-gallery}} and the potential it will have in the coming decade upon the start of its operations. We expect the Legacy Survey of Space and Time (LSST) to discover tens of millions of AGNs - a recent study of QSO number counts puts this to $\sim$12.2 million in i-band stretching to a 5$\sigma$ median depth of 26.4 mag (\citealt{LSST2019ApJ}, Li et al., under review). The latest survey simulations (v5.0) suggest a mean cadence of nearly every half-night in \textit{riz} bands in the Wide-Fast-Deep (WFD) mode, while this can be as high as every 3 nights for the u-band. The numbers are slightly more optimistic for the Deep-Drilling-Fields (DDFs, see e.g., \citealt{2024RNAAS...8...47P}). The sheer amount of data that is going to be involved is nothing that we have analyzed before. Not to forget, it is going to be a daunting challenge to get a spectroscopic follow-up for all these detections, although some upcoming surveys seem promising (ESO/4MOST - \citealt{4MOST2019Msngr.175....3D, Frohmaier2025arXiv250116311F}, Spec-S5 - \citealt{Besuner_2025arXiv250307923B}, WST - \citealt{WST_2024arXiv240305398M}). In anticipation of the start of Rubin operations and to tackle the prospective challenge of spectra-starving, we looked into the large repository of photometrically detected and monitored AGNs in the Zwicky Transient Facility \citep[ZTF,][]{Bellm_2019PASP..131a8002B, Graham_2019PASP..131g8001G} that led to a rather convincing discovery.

\subsection{Photometric variability anti-correlates with the Eddington ratio}

The paper \citep{Benati-Goncalves2025ApJ} presents a systematic study of optical variability for 915 type‑1 quasars with 0 $\leq$ z $\leq$ 3, using ZTF g‑band light curves and Sloan Digital Sky Survey (SDSS) DR16 spectral parameters (including black‑hole masses, bolometric luminosities, and Eddington ratios). These sources were pre-selected from the All Quasar Multiepoch Spectroscopy (AQMES) MEDIUM subsample of the Black Hole Mapper project of the SDSS-V \citep{SDSS5_2019BAAS...51g.274K}. We computed the fractional variability amplitude (F$_{\rm var}$) via the excess‑variance formalism, and after rigorous filtering ($\geq$ 100 observations, F$_{\rm var}>$ 0) and correcting for minimal emission‑line contributions (a few percent of the continuum flux), we arrive at a key result - a redshift‑independent anti-correlation between F$_{\rm var}$ and the Eddington ratio ($\lambda_{\rm Edd}$). The best‑fit relation for the full sample is  

\begin{equation}
\log{\rm \lambda_{Edd}} = (-0.71\pm0.06)\log({\rm F_{\rm var}}) - (1.52\pm0.06) 
\label{eq1}
\end{equation}

\begin{figure}
    \centering
    \includegraphics[width=\linewidth]{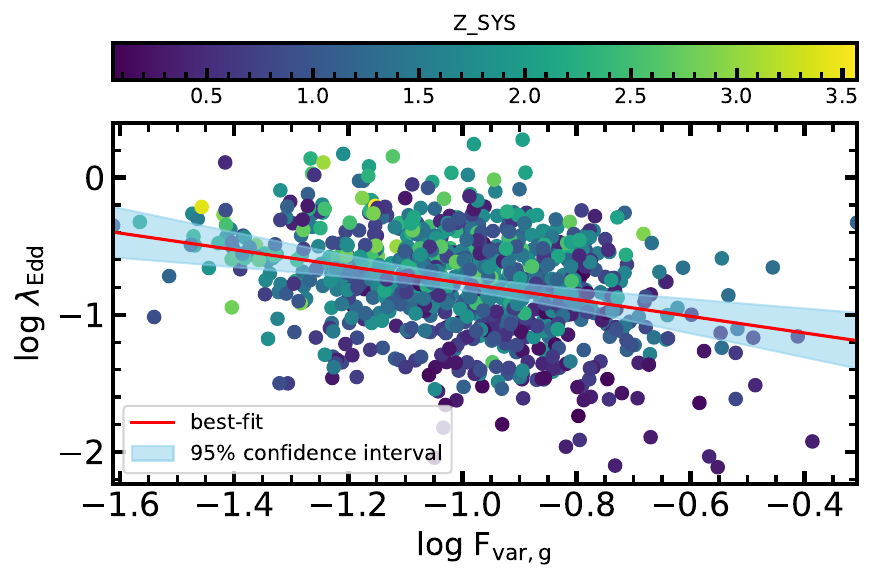}
    \caption{Distribution of the fractional variability (F$_{\rm var}$) in the g-band ZTF lightcurves for the AQMES medium field monitored within the SDSS-V, versus the Eddington ratio. The latter is taken from the SDSS DR16 QSO catalogue \citep{Wu_Shen_2022ApJS..263...42W}. The color axis depicts the distribution of redshift. The best-fit correlation after cleaning sources with insufficient F$_{\rm var}$ in g-band information: log $\lambda_{\rm Edd}$ = -0.61 log F$_{\rm var}$ - 1.38 ($\rho$ = -0.28; p-value = 2.8E-18).}
    \label{fig1}
\end{figure}

or equivalently F$_{\rm var}\approx\,10^{-0.71}\lambda_{\rm Edd}^{-1.52}$ (see, e.g., Figure \ref{fig1} for a revised version). This yields a Pearson coefficient r $\approx$ ‑0.31 (p $<$ 0.01) for all redshifts, with the strongest anti-correlation at low-z (r $\approx$ ‑0.40) and a marginal trend in the highest bin (2 $\leq$ z $<$ 3). We present a general equation encapsulating this relationship, which appears to be almost free of redshift dependence, enabling predictions of quasar variability based on accretion parameters or vice versa, a significant enhancement to prior works \citep[see e.g.,][]{2004ApJ...609...69K}. The derived relation with the Eddington ratio provides a unified framework for interpreting variability in AGNs and facilitates future studies of quasar variability using high-cadence surveys, such as the Vera C. Rubin Observatory’s LSST \citep{LSST2019ApJ}.

Additional correlations show that F$_{\rm var}$ declines monotonically with continuum luminosities L$_{1350}$, L$_{3000}$, L$_{5100}$, and bolometric luminosity across all redshift bins (negative Pearson r values ranging from ‑0.23 to -0.59), while, the F$_{\rm var}$–M$_{\rm BH}$ relationship evolves from a weak positive correlation at low-z (r $\approx$ 0.22) to a moderate anti-correlation at high-z (r $\approx$ ‑0.36). We discussed potential selection biases (luminosity‑driven Malmquist bias, rest‑frame time‑baseline shortening at high-z) but demonstrate that the correlation between F$_{\rm var}$ and $\lambda_{\rm Edd}$ is largely free of any significant redshift dependence \citep[see][for more details]{Benati-Goncalves2025ApJ}. 

\subsection{Possible biases and streamlining the parent sample}

Below, we summarize some additional tests of robustness for our newfound scaling relation.
\begin{itemize}
    \item lightcurve baseline testing: as it currently stands, the correlation involves all AGNs with at least 100 observations in overa  6-year period (mid-2017 to late-2023). The median of the number of g-band observations is 304; thus, we attempted to check the correlation for only those cases where the number of visits was $\geq$ 300. We were left with 53\% of the number of AGNs (488/920). However, the correlation grew stronger: log $\lambda_{\rm Edd}$ = -0.71 log F$_{\rm var}$ - 1.50 ($\rho$ = -0.31; p-value = 2.74E-12), almost identical to the original version as shown in Eq. \ref{eq1}.
    
    \item removing host-dominated sources: Given the wide range of redshift considered in the sample selection (z $\leq$ 3), we decided to also check the variety in our sample with respect to the contamination induced by the presence of a strong host component. This is especially an issue for the low-z sample (z $\leq$ 1), where the fitting routine in the spectral decomposition code PyQSOFit \citep{2018ascl.soft09008G} - the mainstay pipeline used to derive the SDSS DR16 QSO catalog chooses whether to fit a host-galaxy component to the spectrum or not, based on a minimum number of host-galaxy pixels identified by the routine. If this criterion is not met, no host galaxy component is fit. Another way to look at this issue is that if the AGN dominates the spectrum, the resulting fit will have a negligible host-galaxy contribution. At the moment of writing, we do not have a straightforward way to filter such sources out without this information. On the other hand, for the sources with z $>$ 1, no host galaxy template is incorporated (the host galaxy template fits within the rest-frame SDSS range, i.e., 3450-8000 \AA). Therefore, we only rely on the Malmquist bias for this latter case. Nonetheless, we made a simple test to account for the host contribution. The QSO catalog does provide us with the fractional host contribution at 5100\AA\ (${\rm f_{host, 5100}}$). We plotted the distribution for ${\rm f_{host, 5100}}$ and noted a clear dichotomy at 0.4 (see Figure \ref{fig2} left panel). After filtering out the sources' significant host contribution ($>$ 40\%), resulting in recovery of 86\% of the sample (792/920), we still recover a familiar version of the correlation between $\lambda_{\rm Edd}$ and F$_{\rm var}$.

    \begin{figure}
        \centering
        \includegraphics[width=0.495\linewidth, height=0.35\linewidth]{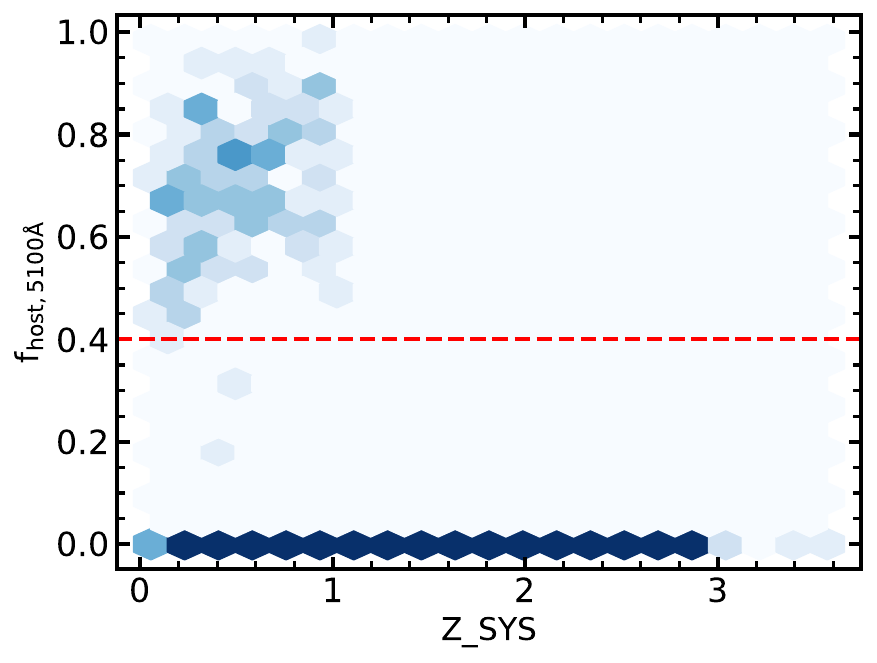}
        \includegraphics[width=0.495\linewidth, height=0.36\linewidth]{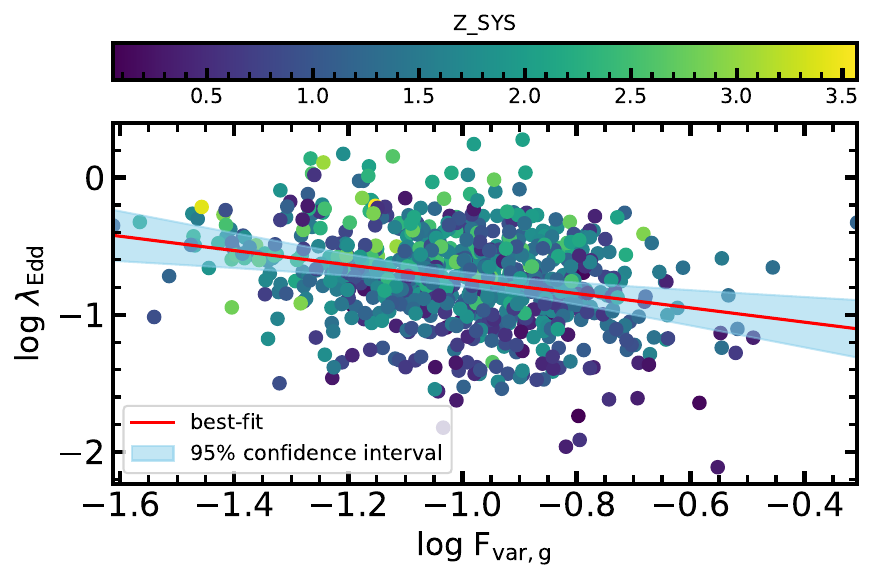}
        \caption{(Left:) Host fraction (estimated at 5100 \AA) as a function of redshift for our sample. The hexagons highlight the number density of sources with a darker shade of blue corresponding to higher density. (Right:) The best-fit correlation after filtering out the sources with significant host contribution ($>$40\%): log $\lambda_{\rm Edd}$ = -0.52 log F$_{\rm var}$ - 1.26 ($\rho$ = -0.27; p-value = 7.1E-15).}
        \label{fig2}
    \end{figure}

    \item combining the above two cases: taking the two above cases together and making the cut, we recover a correlation similar to the original one (with 44\% of the sample): log $\lambda_{\rm Edd}$ = -0.62 log F$_{\rm var}$ - 1.37 ($\rho$ = -0.3; p-value = 9.78E-10).

    \item quality cut on the relative error in F$_{\rm var}$: Next we capitalize on the relative error in the F$_{\rm var}$ itself. The errF$_{\rm var}$/F$_{\rm var}$ in the g-band has a distribution with a median around 57\%. Thus, we tried a simple cut to limit to the sources with less than 50\% in relative error in F$_{\rm var}$. This led to a substantial reduction in the remaining sources ($\sim$39\%; 357/920). Applying this cut only, we get a stronger correlation than its original form: log $\lambda_{\rm Edd}$ = -0.83 log F$_{\rm var}$ - 1.54 ($\rho$ = -0.32; p-value = 5.78E-10).

    \begin{figure}
        \centering
        \includegraphics[width=\linewidth]{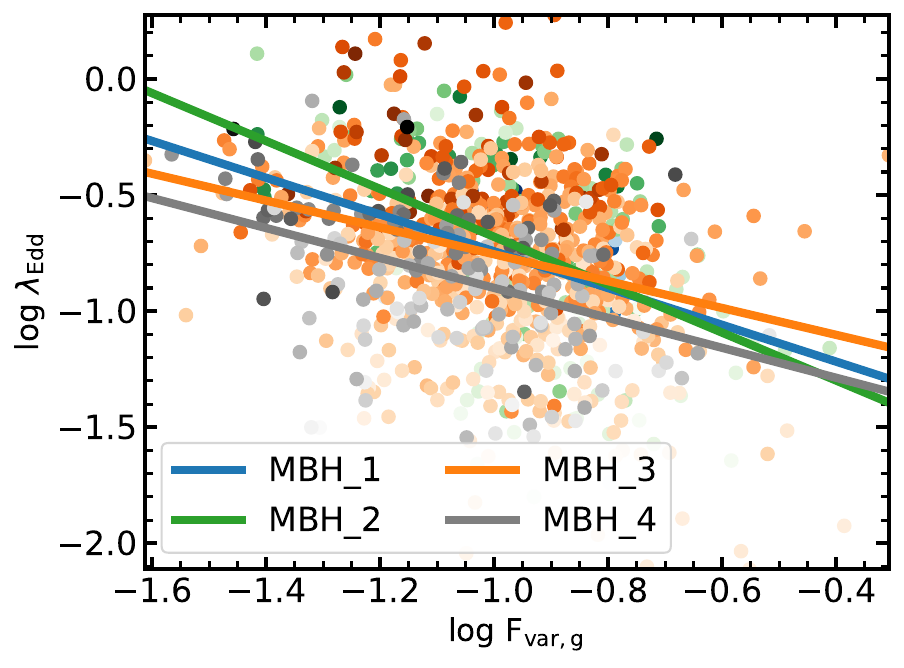}
        \includegraphics[width=0.495\linewidth, height=0.3225\linewidth]{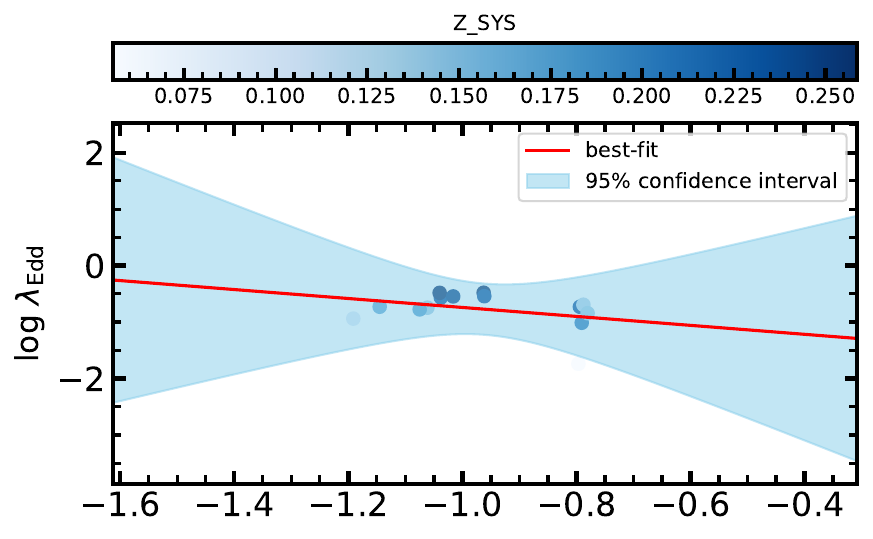}
        \includegraphics[width=0.495\linewidth]{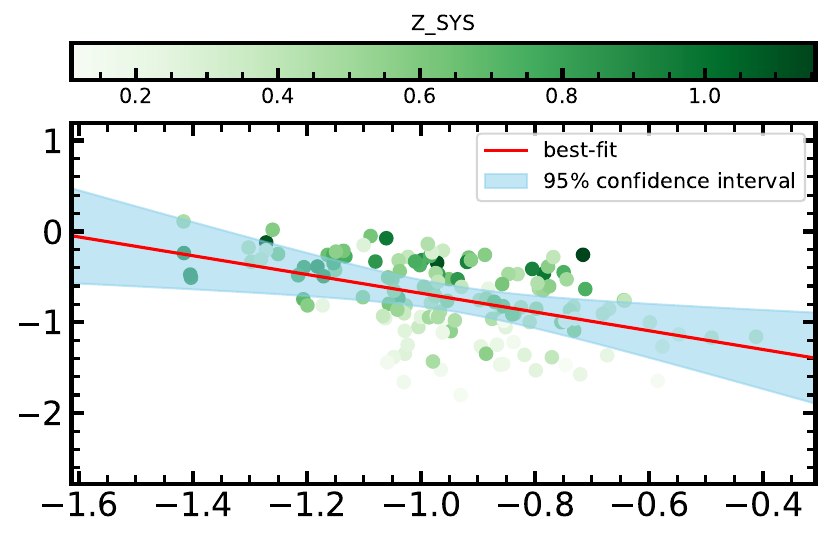}
        \includegraphics[width=0.495\linewidth, height=0.3475\linewidth]{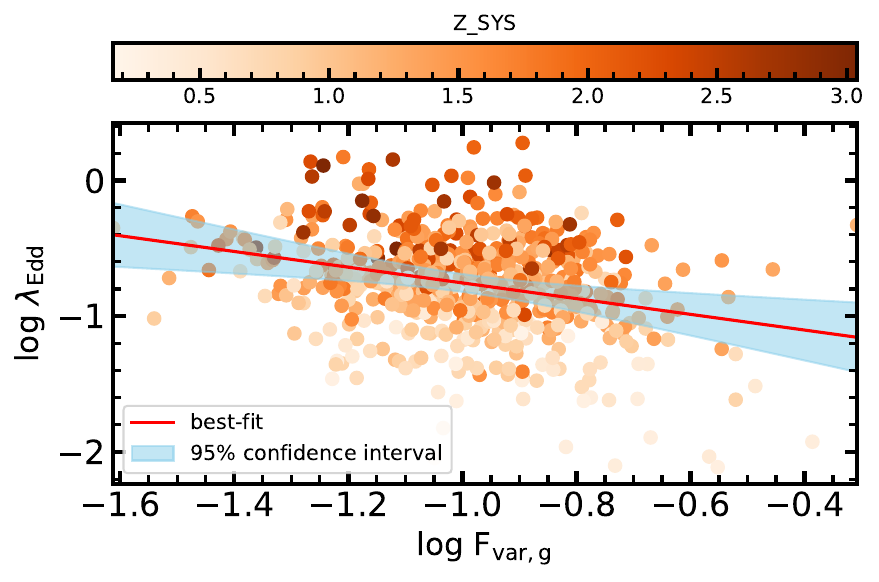}
        \includegraphics[width=0.495\linewidth]{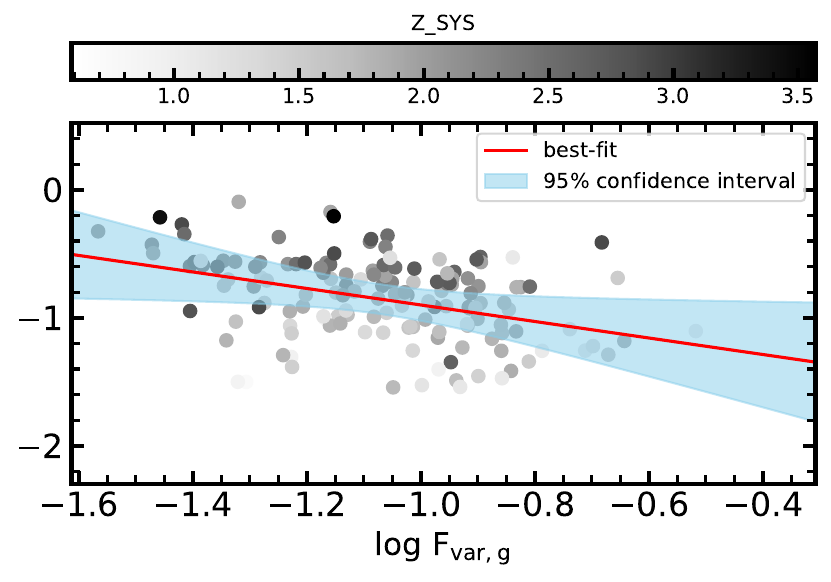}
        \caption{Correlations similar to as shown in Figure \ref{fig1}, but with the data binned in four separate bins of M$_{\rm BH}$ (in log-scale, in M$_{\odot}$ units): between 6.5 - 7.5 (blue), 7.5 - 8.5 (green), 8.5 - 9.5 (orange), and 9.5 - 10.5 (grey). The sources are colored as per their respective M$_{\rm BH}$ bins, while the gradient in color represents the corresponding range in redshift as depicted in the lower sub-panels. The best-fit per M$_{\rm BH}$ bin is shown in the top panel. Each of the lower sub-panels shows the individual best-fit to the binned data per M$_{\rm BH}$ bin. The color-axis in each of these sub-panels is the redshift corresponding to the respective bin. Note the shifting range in the redshift range as we move along the M$_{\rm BH}$ bins.}
        \label{fig:mbh}
    \end{figure}

    \item dependence on the M$_{\rm BH}$ bins: To check the co-dependence of M$_{\rm BH}$ on the derived correlation, we made a test by binning the sources within 4 equal bins of M$_{\rm BH}$. The M$_{\rm BH}$ distribution for the cleaned sample of 920 sources ranges from 10$^{6.81}$ to 10$^{10.32}$ M$_{\odot}$, with a mean $\sim$10$^{8.99}$ M$_{\odot}$ ($\sigma$ = 0.55 dex; median mass = $\sim$10$^{9.09}$ M$_{\odot}$). We thus constructed 4 bins, each of 1 dex width, i.e., between 6.5 - 7.5, 7.5 - 8.5, 8.5 - 9.5, and 9.5 - 10.5. These bins have 14, 148, 607, and 151 sources, respectively. The correlation for each M$_{\rm BH}$ bin is found as follows:
    \begin{itemize}
        \item bin-1: log $\lambda_{\rm Edd}$ = -0.79 log F$_{\rm var}$ - 1.53 ($\rho$ = -0.1; p-value = 0.725)
        \item bin-2: log $\lambda_{\rm Edd}$ = -1.03 log F$_{\rm var}$ - 1.71 ($\rho$ = -0.47; p-value = 2.32E-09)
        \item bin-3: log $\lambda_{\rm Edd}$ = -0.58 log F$_{\rm var}$ - 1.33 ($\rho$ = -0.25; p-value = 7.97E-10)
        \item bin-4: log $\lambda_{\rm Edd}$ = -0.65 log F$_{\rm var}$ - 1.54 ($\rho$ = -0.38; p-value = 1.62E-06)
    \end{itemize}
    Given that most of the sources (66\%) are in bin-3, the resemblance of the correlation for this bin relative to the cleaned sample is expected. We note also the shifting range in the redshift range as we move along the M$_{\rm BH}$ bins, which is connected to the Malmquist bias (wherein we tend to observe the brighter targets as we transcend in redshift). Another reason is the sensitivity of our spectrograph detectors, which marks a limiting magnitude below which the detection of targets is seldom. Given the direct proportionality between the M$_{\rm BH}$ and luminosity, the faintness of the target could lead to the estimation of a lower M$_{\rm BH}$ for these sources. We, however, miss these sources as we go along the redshift. We, therefore, need better facilities (e.g., DESI; \citealt{Pucha2025ApJ...982...10P}) to probe this low-luminosity regime, esp. in the higher z regime, to have a better handle on completeness of the sample. 
    
    However, interestingly, the correlation for bin-2 (7.5 - 8.5) is found to be the strongest among all ($\rho$ = -0.47 with a slope suggesting an almost 1-to-1 inverse relation between F$_{\rm var}$ and $\lambda_{\rm Edd}$).  The top panel of Figure \ref{fig:mbh} shows the consolidated version of each bin's performance. This reveals an interesting conclusion: the dispersion between the individual best-fits varies between $\sim$ 0.5 dex (for the smallest values of log F$_{\rm var}$) to $\sim$ 0.25 dex (for the largest values of log F$_{\rm var}$). These values are consistent with the dispersion in the M$_{\rm BH}$ scaling relations often adopted (e.g., the M-$\sigma_{\star}$ relation with a dispersion $\sim$0.44 dex, \citealt{Prieto_2022MNRAS.510.1010P}; and similarly for RM-derived M$_{\rm BH}$, i.e., between 0.4 - 0.51 dex, \citealt{Du_Wang_2019}). Also, it is interesting to see that, apart from the M$_{\rm BH}$ bin containing the largest BH massed sources, the rest of the three relations coincide around log F$_{\rm var}$ $\sim$ -0.9 and log $\lambda_{\rm Edd}$ $\sim$ -0.75. This could suggest that the correlation briefly becomes independent of the M$_{\rm BH}$ variations within this region.
\end{itemize}

Overall, these experiments only reinforce the \textit{almost} universality of the relation found between the fractional variability (F$_{\rm var}$) and the Eddington ratio ($\lambda_{\rm Edd}$). The derived universal F$_{\rm var}$–$\lambda_{\rm Edd}$ relation, therefore, enables predictions of quasar variability from accretion parameters and vice versa, offering a powerful tool for upcoming high‑cadence surveys.

\section{Changing-look and how?}
\label{sec2}

The last decade has seen a surge in the detections of changing-look AGNs (CLAGNs) - these sources have marked increases in their variability features compared to their counterparts, which we discussed in the previous section; thanks to massive, multiplex spectroscopic surveys such as SDSS and DESI, to name a few. Combination of intensive photometric monitoring, e.g., Catalina Real-Time Transient Survey \citep[CRTS,][]{CRTS_2009ApJ...696..870D} and Zwicky Transient Facility \citep[ZTF,][]{Bellm_2019PASP..131a8002B, Graham_2019PASP..131g8001G} and multi-epoch spectroscopy have revealed $\geq$ 500 known CLAGNs to date \citep{Temple_2023MNRAS.518.2938T, Wang_2024ApJ...966..128W, Zeltyn_2024ApJ...966...85Z, Guo_DESI_2025ApJS..278...28G}. Many of these detections were serendipitous to start with, but as we prepare for Rubin LSST to start its operations, the community has become more aware of selection criteria to look for CLAGNs in massive datasets, and even forecast the changing-state of these AGNs \citep{Graham_2020MNRAS.491.4925G, Sanchez-Saez_2021AJ....162..206S, Ricci_2023NatAs...7.1282R}. With the aid of beyond-meter class facilities, e.g., Gemini/ESO-VLT/DESI, we now peer deep into the redshift space to reveal CLAGN candidates close to the cosmic noon \citep{Ross_2020MNRAS.498.2339R, Guo2_2025ApJ...981L...8G}. Understanding the nature of this special class of AGNs in contrast to the general variable nature of AGNs is vital to our understanding and perhaps aids in revealing new classes of CLAGNs, e.g., the extreme variability quasars \citep[EVQs,][]{2022ApJ...925...50R}.\\

\begin{figure}
    \centering
    \includegraphics[width=0.495\linewidth]{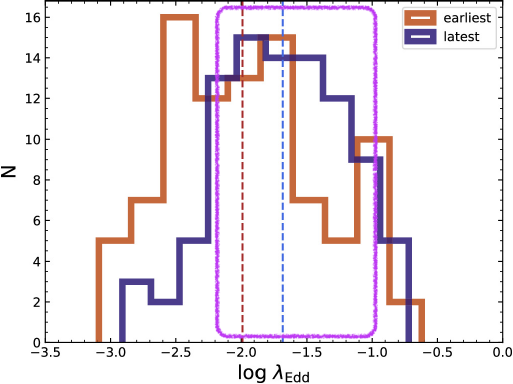}
    \includegraphics[width=0.495\linewidth]{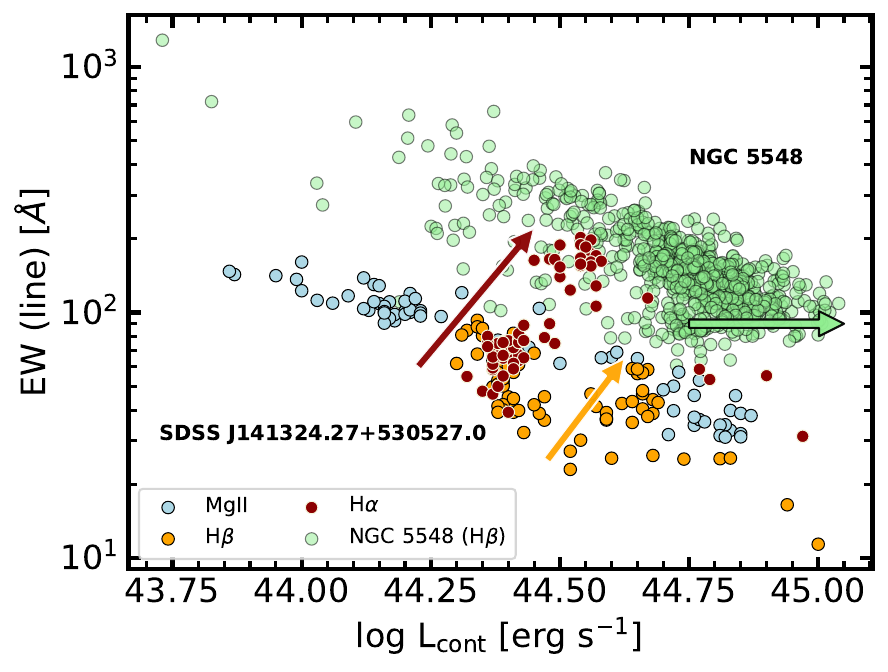}
    \caption{(\textit{Left:}) Distribution of Eddington ratios in the sample from \citet{PandaSniegowska2024ApJS}. We show the distributions for the earliest (in brown) and the latest (in purple) epochs for the sources in our sample. The median values for the two distributions (red = -1.99, blue = -1.685) are marked with vertical dashed lines. The magenta box marks the range of the Eddington ratio for NGC 5548, i.e., log $\lambda_{\rm Edd}$ = [-2.2, -1]. (\textit{Right:}) The distribution of the emission line EW versus the AGN continuum luminosity. Here, we demonstrate the trend for two sources:  SDSS J141324.27+530527.0 \citep{Wang_2018ApJ...858...49W} with 72 spectral epochs over $\sim$15 years (5527 days), and NGC 5548 \citep{Bon_2018FrASS...5....3B, Panda2022AN, Panda2023BASBr} with more than 750 spectral epochs over $\sim$25 years (9624 days). For the former source (SDSS J141324.27+530527.0), we have taken the spectral data from the homogeneous fitting in \citet{PandaSniegowska2024ApJS}, which includes the Mg{\sc ii}, H$\beta$, and H$\alpha$ emission lines and the corresponding AGN continua nearest to these lines (at 3000\AA, 5100\AA, and 6000\AA), as shown in the panel. For NGC 5548, we show the dataset from Panda et al. (in prep.), which is an updated version of the dataset provided in \citet{Bon_2018FrASS...5....3B}, where the authors analyzed the H$\beta$ region. The SDSS source shows a clear rise from a deep minimum to a high state in both Balmer lines. However, the Mg{\sc ii} shows a rather flat behavior - reminiscent of the Baldwin effect, suggesting the difference in ionization and response to the changing continuum levels. NGC 5548 data has a wealth of data, but the change in the source is rather gradual, and hence, a clear spike in the trend is not that prominent.}
    \label{fig:clagns}
\end{figure}

In this direction, we \citep{PandaSniegowska2024ApJS} made a significant stride to reveal where the population of the CLAGNs is mostly discovered through a systematic analysis of known SDSS CLAGNs. In the next sub-section, we describe our methodology and the key findings that connect again the key parameter - Eddington ratio (see left panel in Figure \ref{fig:clagns}).

\subsection{Eddington ratio distribution as a proxy to catch Changing-Look AGNs} 

The paper by \citet{PandaSniegowska2024ApJS} presents a homogeneous spectroscopic analysis of 93 CLAGNs compiled from the SDSS, BOSS, and eBOSS archives. Multi‑epoch spectra are processed with the PyQSOFit pipeline \citep{2018ascl.soft09008G}, delivering epoch‑resolved measurements of the AGN continuum, broad‑line parameters, black‑hole mass (M$_{\rm BH}$), and Eddington ratio ($\lambda_{\rm Edd}$). The sample was assembled by cross‑matching existing CLAGN catalogs and retrieving all available SDSS spectra, ensuring at least two epochs with detectable broad H$\beta$ emission profiles. Spectral decomposition follows the standard PyQSOFit methodology, incorporating a power‑law continuum, Fe {\sc ii} emission templates, and host‑galaxy eigenspectra when required. Each epoch is placed on the optical Eigenvector 1 (EV1) plane, i.e, the optical plane between the broad FWHM(H$\beta$) versus the strength of the optical Fe {\sc ii} emission, i.e., the parameter R$_{\rm Fe}$. This is the canonical quasar main‑sequence diagram \citep{1992ApJS...80..109B, Sulentic_2000, 2001ApJ...558..553M, Marziani_2018, Panda2019_Orientation, Panda2024FrASS}. The majority of CL-AGNs evolve within Population-B (FWHM H$\beta$ $>$ 4000 km/s), with only a few exhibiting inter‑population transitions (A $\rightleftharpoons$ B). Turn‑on events are accompanied by systematic shifts toward lower M$_{\rm BH}$ and, therefore, higher $\lambda_{\rm Edd}$, supporting accretion‑rate modulation as the primary driver of the phenomenon. We then classified the sources as ``Turn‑On'', ``Turn‑Off'' or mixed ``On‑Off'' cycles; and found that the distribution is dominated by simple monotonic transitions, echoing earlier reports that turn‑off events are more frequently observed \citep{Shen_2021ApJ...918L..19S}. Balmer‑decrement (H$\alpha$/H$\beta$) variability is examined for 32 objects with at least 3 epochs. The ratios display diverse temporal behavior, often deviating from the case-B value ($\sim$3.1, \citealt{Osterbrock_Ferland_2006}), implying a complex interplay among dust extinction, BLR density, and ionizing‑continuum changes. The resulting database furnishes a comprehensive resource for probing the physical mechanisms behind AGN spectral variability. The systematic migration of CLAGNs on the EV1 and M$_{\rm BH}$ - $\lambda_{\rm Edd}$ planes bolsters models in which rapid accretion‑rate fluctuations dominate the changing‑look phenomenon \citep{2018MNRAS.480.3898N, 2020A&A...641A.167S}, while heterogeneous Balmer‑decrement trends suggest additional BLR structural evolution.

The main finding from this study was the realization of the Eddington ratio distribution of these CLAGNs. We specifically consider the spectral epochs as widely separated in time and with the most variation in the continuum and emission line characteristics to reveal that the CLAGNs dominate the low-accretion regime ($\lambda_{\rm Edd} \sim$ 0.01), dominated by sources in the Population B (FWHM H$\beta$ $\geq$ 4000 km s$^{-1}$), in contrast to the high accretors that dominate the Population A parameter space \citep{Du_2018, Panda2019_Orientation, 2021ApJ...910..115S, Garnica_2022A&A...667A.105G}.

\subsection{Population studies vs. tracking the changes in an individual source}

Going back to the BLR structural evolution, it would be great if we could capitalize on single-object studies over longer temporal baselines to reveal and distinguish between short, intermediate, and long-term variations in these AGNs. We are fortunate to have such rich datasets focused on single AGN spectrophotometric monitoring, notably of the prototypical Population B source - NGC 5548. Taking advantage of $\geq$ 30 years of data compiled from $\sim$ 17 individual observing campaigns - including the AGN Watch and AGN STORM projects \citep{Peterson_2002ApJ...581..197P, Shapovalova_2004A&A...422..925S, DeRosa_2015ApJ...806..128D, Fausnaugh_2016ApJ...821...56F, Pei_2017ApJ...837..131P}, we \citep{Panda2022AN, Panda2023BASBr} re-discovered the saturation of the H$\beta$ emission with growing AGN continuum luminosity, also known as the Pronik-Chuvaev effect \citep{Pronik_1972Ap......8..112P}. NGC 5548 has telltale signatures of being a CLAGN, and this anomalous behavior of the H$\beta$ with changing continuum luminosity  \citep{Gaskell_2021MNRAS.508.6077G} has opened new avenues for us to improve our spectral models to not only reveal the inner working of the AGN, but to use this source as a laboratory to better understand the CLAGN behavior. For instance, in the right panel of Figure \ref{fig:clagns}, we demonstrate the trend between the EW(H$\beta$) with increasing AGN luminosity, the well-known Baldwin effect \citep{1999ApJ...527..649W, MLMA_2021} - however, we see a flattened behavior at the extreme high end of the luminosity. This is the Pronik-Chuvaev effect. We try to check whether the CLAGNs that were studied in \citet{PandaSniegowska2024ApJS} show a similar behavior to NGC 5548. To this end, we consider our best case - SDSS J141324.27+530527.0 \citep{Wang_2018ApJ...858...49W}, an object with 72 spectral epochs spanning over $\sim$15 years (5527 days). Another interesting thing with the latter source is the availability of EW information for Mg {\sc ii}, and H$\alpha$, in addition to H$\beta$ for all epochs. We overlay the EW for these three emission lines in the same figure as NGC 5548. To be consistent, we utilize the continuum closest to each line. For Mg {\sc ii} this continuum was extracted at 3000\AA\, while for H$\beta$ and H$\alpha$, we use the continua at 5100\AA\ and 6000\AA\, respectively. We see a clear flattening behavior in H$\beta$ in the SDSS source at higher luminosities; however, we note a rise-and-fall effect in the Balmer lines - H$\alpha$ rises first, followed by H$\beta$. Whereas, the Mg {\sc ii} doesn't show any such prominent change along the luminosity trend. Not so surprisingly, the SDSS source, during its 72 epochs, modulated in Eddington ratio between (in log-scale) -0.75 to -2.5. This range is not so different from NGC 5548, which has been observed to vary in Eddington ratio between (in log-scale) -1 to -2.2. Is this a coincidence, or are we onto something? This definitely needs further investigation.

Looking into the future, we are now well-poised to utilize these findings to pre-select changing-look (and changing-state) candidates in large datasets and up-and-coming high-fidelity surveys, in conjunction with variability statistics, color-color diagrams, and marked flux variations \citep{Shen_2021ApJ...918L..19S, Ricci_2023NatAs...7.1282R}.

\section{Resolving and revealing: highlights from SOAR/SIFS}
\label{sec3}

\begin{figure}
    \centering
    \includegraphics[width=\linewidth]{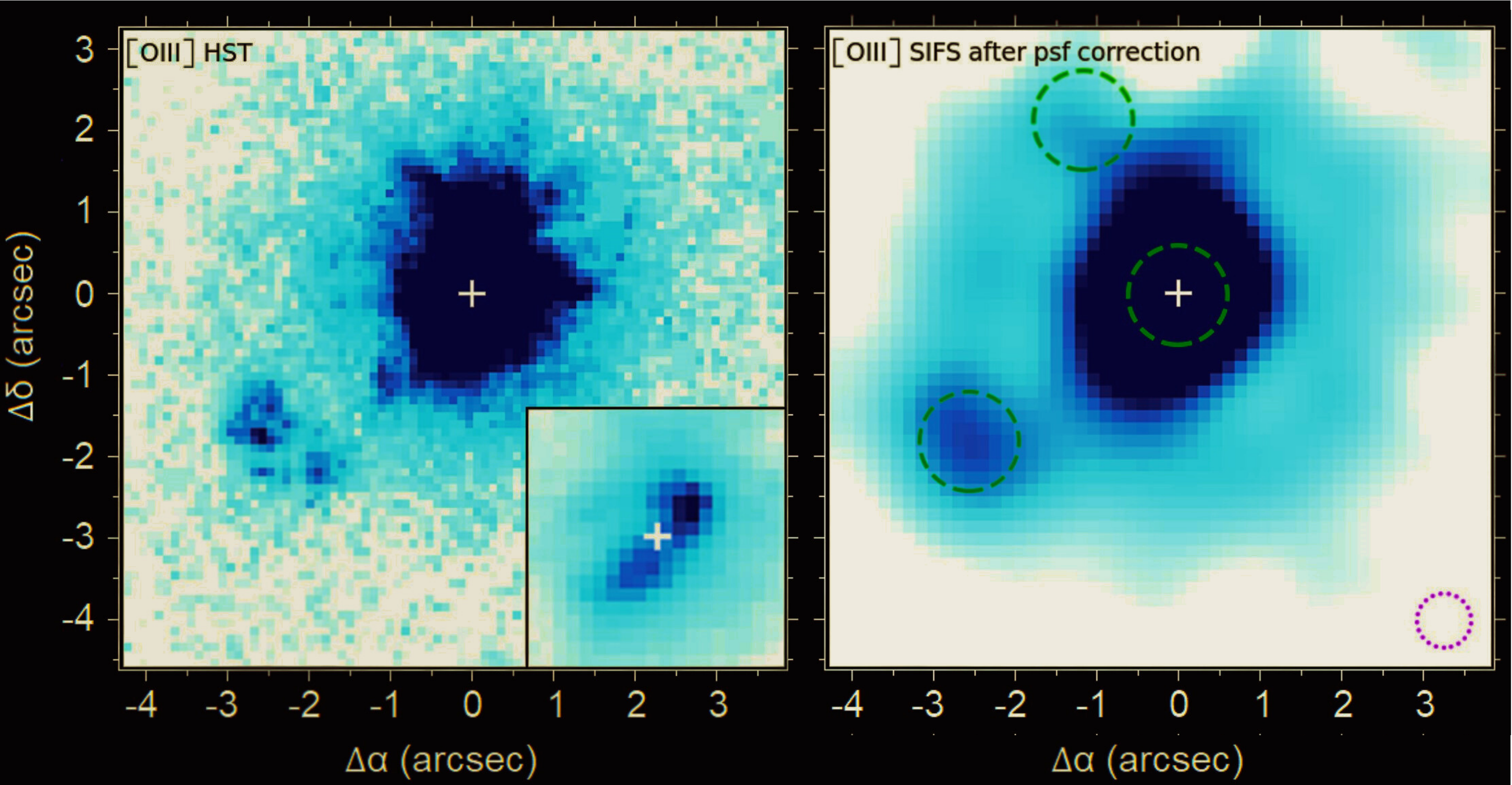}
    \caption{(\textit{Left:})[O {\sc iii}]$\lambda$5007 emission image of ESO 138-G001 from the HST/WFPC2 \citep{HST2000ApJS..128..139F} with a small inset showing the details of the central part, where the nucleus is marked with a plus sign; (\textit{Right:}) [O {\sc iii}]$\lambda$5007 emission image from SIFS after data treatment involving spatial re-sampling with quadratic interpolation followed by the Richardson-Lucy PSF deconvolution \citep[see][for more details]{SIFS2024MNRAS}. The green circles – with an aperture radius of 0.6 arcsec – denote the extraction regions of the spectra for the North-East (NE) knot (top left), the South-East (SE) blob (bottom left), and the nuclear region. The red circle denotes the PSF FWHM of 0.71 arcsec.}
    \label{fig:hst-sifs}
\end{figure}

Continuing our venture in the low-accretion regime, AGN spectral energy distributions (SEDs) provide a powerful diagnostic to probe the dominant ionization mechanism, test against high-quality observations, and assess the true accretion state of the AGN. In this context, we present a salient case that combines state-of-the-art photoionization modeling with one of the first high-angular-resolution integral field unit (IFU) observations obtained with the 4m SOAR telescope. This study, focused on the nearby Seyfert 2 galaxy ESO 138-G001, directly contests earlier interpretations based on HST imaging (see Figure \ref{fig:hst-sifs}). We confirm that the observed spectra are unequivocally AGN-dominated and further demonstrate that peripheral emission structures arise from filtered radiation originating from the nucleus. Using a quantitative, iterative technique, we successfully generated synthetic spectra in remarkable agreement with both nuclear and off-nuclear observations, reinforcing the AGN-driven photoionization of this low-accreting Type-2 AGN extending beyond a few hundred parsecs from the nucleus.\\

\begin{figure}
    \centering
    \includegraphics[width=0.475\linewidth]{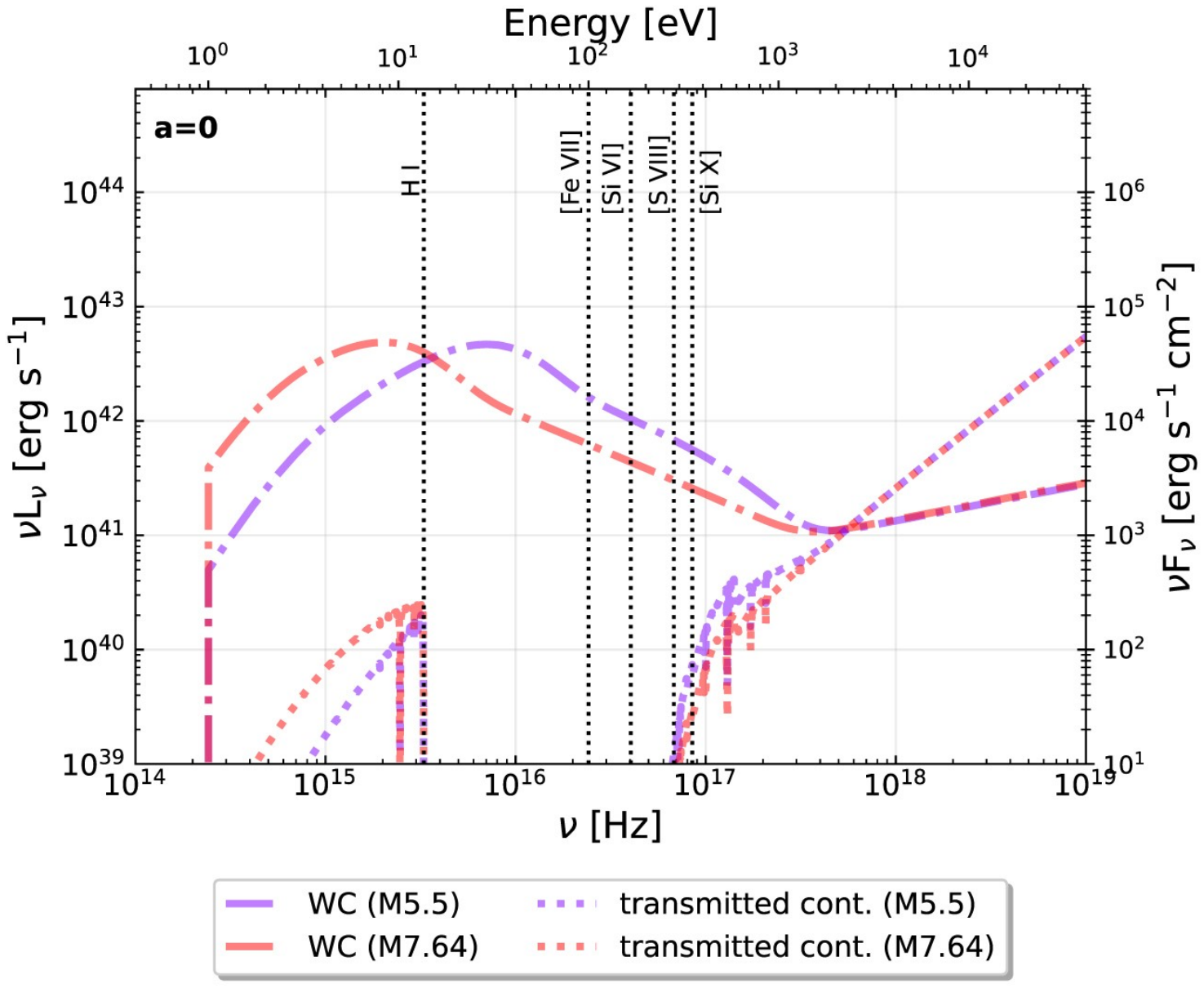}
    \includegraphics[width=0.45\linewidth]{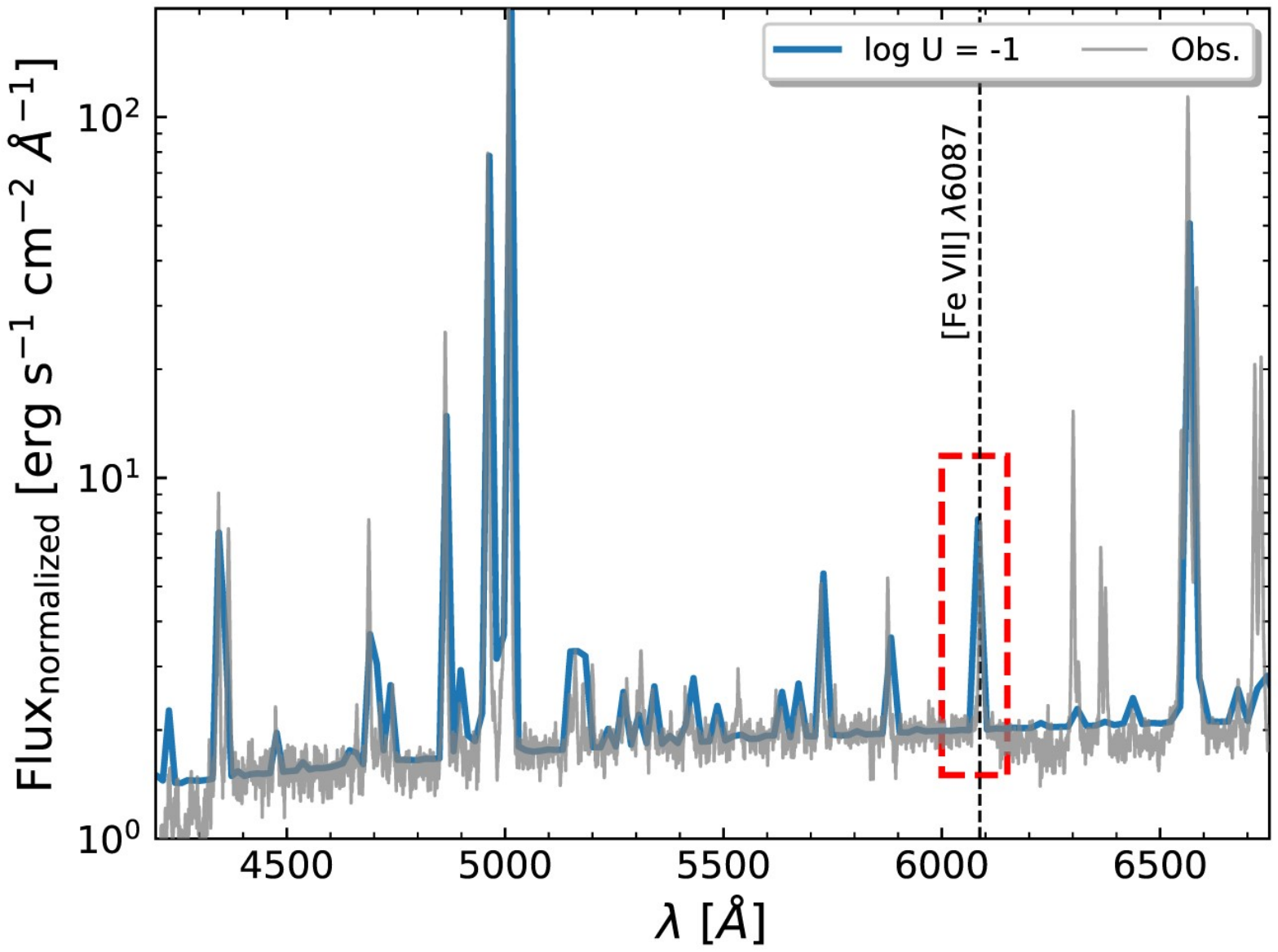}
    \includegraphics[width=\linewidth]{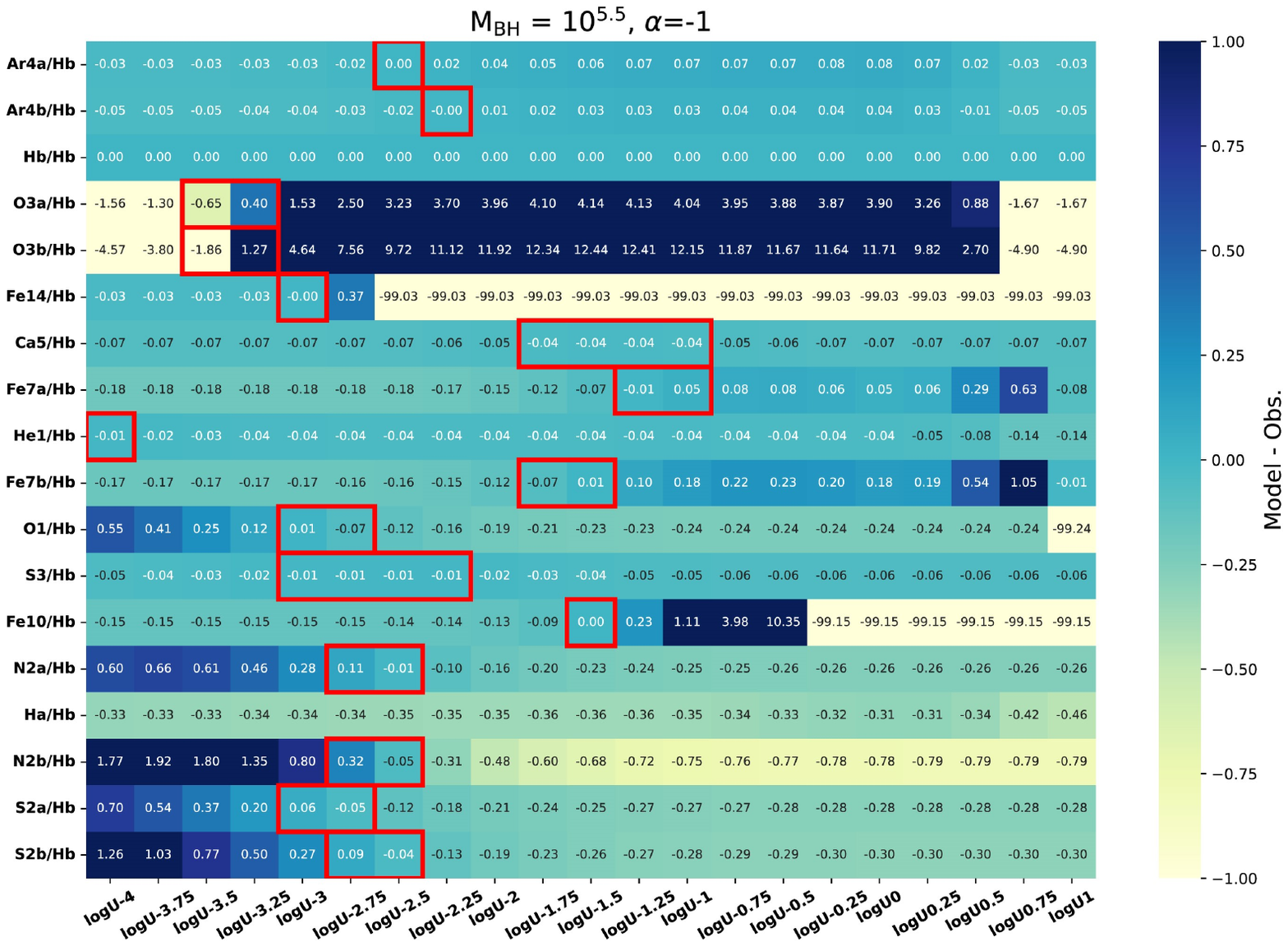}
    \caption{(\textit{Top left:}) Incident SEDs generated for our photoionization modeling. The incident continua for the two black hole mass cases are shown in dot–dashed. These are used for the modeling of the nuclear region. The corresponding transmitted continua from these models are shown in dotted lines. These latter distributions are used as incident continua for the SE blob. These SEDs have been made assuming a non-spinning black hole. The IPs for notable coronal lines along with the hydrogen ionization front at 13.6 eV are marked with vertical lines; (\textit{Top right:}) Synthetic spectrum (blue) comparison with observed spectrum (in grey), for the nuclear region, generated for the SED corresponding to the M$_{\rm BH}$ = 10$^{5.5}$ M$_{\odot}$, and ionization parameter, log U = -1. (\textit{Bottom:}) Heatmap for the same BH mass case, M$_{\rm BH}$ = 10$^{5.5}$ M$_{\odot}$, for the density law slope, $\alpha$ = -1, for the nuclear region. Each of the considered emission lines is normalized to the H$\beta$ emission line. The x-axis represents the range of ionization parameters considered in our models in each panel. The cases of ionization parameters for each line ratio that have the smallest residuals (modeled ratio–observed ratio) are highlighted with red boxes. Courtesy: \citet{SIFS2024MNRAS}.}
    \label{fig:collage}
\end{figure}

In \citet{SIFS2024MNRAS}, we investigated the inner $\sim$600 pc of ESO 138-G001 using the SOAR Integral Field Spectrograph\footnote{The SIFS IFU employs a lenslet–fiber array with 1300 elements, yielding 1300 simultaneous spectra across a 15 × 7.8 arcsec$^2$ field of view at a spatial sampling of 0.30 arcsec per fiber. The data span 4200–7000 Å with $R \sim 4200$.} (SIFS; \citet{Lepine_2003SPIE.4841.1086L}). ESO 138-G001 is the nearest known Coronal Line Forest (CLiF) AGN \citep{Rose_2015MNRAS.448.2900R}, characterized by unusually strong high-ionization lines (IP $\geq$ 100 eV) compared to typical AGN. This makes it an ideal laboratory to investigate the physics of coronal line emission. Earlier long-slit optical and near-infrared studies by \citet{Cerqueira_2021MNRAS.500.2666C} revealed a hidden broad-line region and a rich set of coronal lines, extending up to [Fe {\sc xiii}] (IP = 330.8 eV) in the NIR.\\

From multiple, strong, narrow emission lines, we obtained $z = 0.00914$, corresponding to a physical scale of $\sim$146 pc arcsec$^{-1}$. ESO 138-G001 is notable for its compact and intense coronal line emission, as well as a bright high-excitation blob $\sim$3 arcsec southeast of the nucleus. The spatial resolution of SIFS enabled us to show that the coronal line forest emission is confined to a $\sim$0.8 arcsec region centered on the nucleus and is powered by the AGN continuum. Radiative transfer modeling further established that the AGN spectrum is filtered by circumnuclear gas within a few tens of parsecs, such that the ionization cone and the SE blob are illuminated by a modified SED dominated by low- to mid-ionization lines, with no evidence for coronal lines (Figure \ref{fig:collage}).\\

Photoionization modeling with {\tt cloudy} \citep{Ferland_2017RMxAA..53..385F} reproduced the observed coronal line spectrum and implied a black hole mass of $\sim$3.2 $\times$ 10$^5$ M$_\odot$, consistent with the value obtained from prior X-ray variability estimates from XMM-Newton \citep{Hernandez_2015A&A...579A..90H}. The inferred accretion rate is $\sim$1\% of the Eddington limit. Our models recovered nearly all observed permitted and forbidden transitions, providing strong validation of the approach.\\

The incident SEDs were generated using the AGNSED three-component framework (disk, warm corona, hot corona; \citealt{Kubota_2018MNRAS.480.1247K}), which has proven effective for modeling coronal line emission across a wide range of AGN properties \citep{Prieto_2022MNRAS.510.1010P}. Applying this framework to ESO 138-G001, we leveraged its known redshift, black hole mass, and Eddington ratio to construct ionizing continua tailored to the source. While archival studies suggested discrepant mass estimates (log M$_{\rm BH}$ = 5.5 vs. 7.64; in units of M$_{\odot}$), our iterative spectral synthesis confirmed the lower value, consistent with the X-ray results, and an accretion rate of $\lambda_{\rm Edd} \approx 0.01$.\\

Kinematic analysis revealed broad Gaussian components in the brightest nuclear lines (e.g., [O {\sc iii}], H$\alpha$, [Fe {\sc vii}]) with FWHM $\sim$450–600 km s$^{-1}$, indicative of a nuclear outflow confined within the central arcsecond. Electron densities span $10^{3-4}$ cm$^{-3}$, reaching $\sim$5.7 $\times$ 10$^3$ cm$^{-3}$ at the nucleus, while [O {\sc iii}] ratios imply $T_e \sim 2.1 \times 10^4$ K. The SE blob, at $\sim$2.6 arcsec from the nucleus, displays only low- to mid-ionization lines and no coronal emission, consistent with photoionization by a filtered AGN continuum. The NE knot, in contrast, shows disturbed kinematics suggestive of interaction with a radio jet.\\

Taken together, these results establish ESO 138-G001 as a prototype CLiF AGN hosting a highly compact coronal line region, a modest nuclear outflow, and an ionization-cone geometry in which the SE blob and NE knot delineate the cone edges. The detailed interplay of coronal line emission, filtered radiation, and outflow/jet dynamics provides one of the clearest demonstrations to date of how AGN SEDs govern line emission in the low-accretion regime.


\section{A faster, efficient way to derive masses of supermassive black holes}
\label{sec4}

Let us shift gears and look into the other end of the redshift/luminosity - the high redshift, high luminosity end, which fittingly brings us to the high-Eddington sources with a relatively small increase in the black hole masses. Reverberation mapping (RM) at high redshift is notoriously time-intensive, often requiring monitoring campaigns that span decades. Landmark studies such as \citet{Lira2018ApJ...865...56L} and \citet{Kaspi2021ApJ...915..129K} demonstrated this challenge, as luminous quasars at $z \sim 2$–3, while ideal for RM due to their high accretion rates and massive black holes, demand prohibitively long temporal baselines. Even with decade-long surveys like LSST, obtaining reliable BLR lags for these sources remains a formidable task.\\

\begin{figure}
    \centering
    \includegraphics[width=0.85\linewidth, trim=0 0 0 14pt, clip]{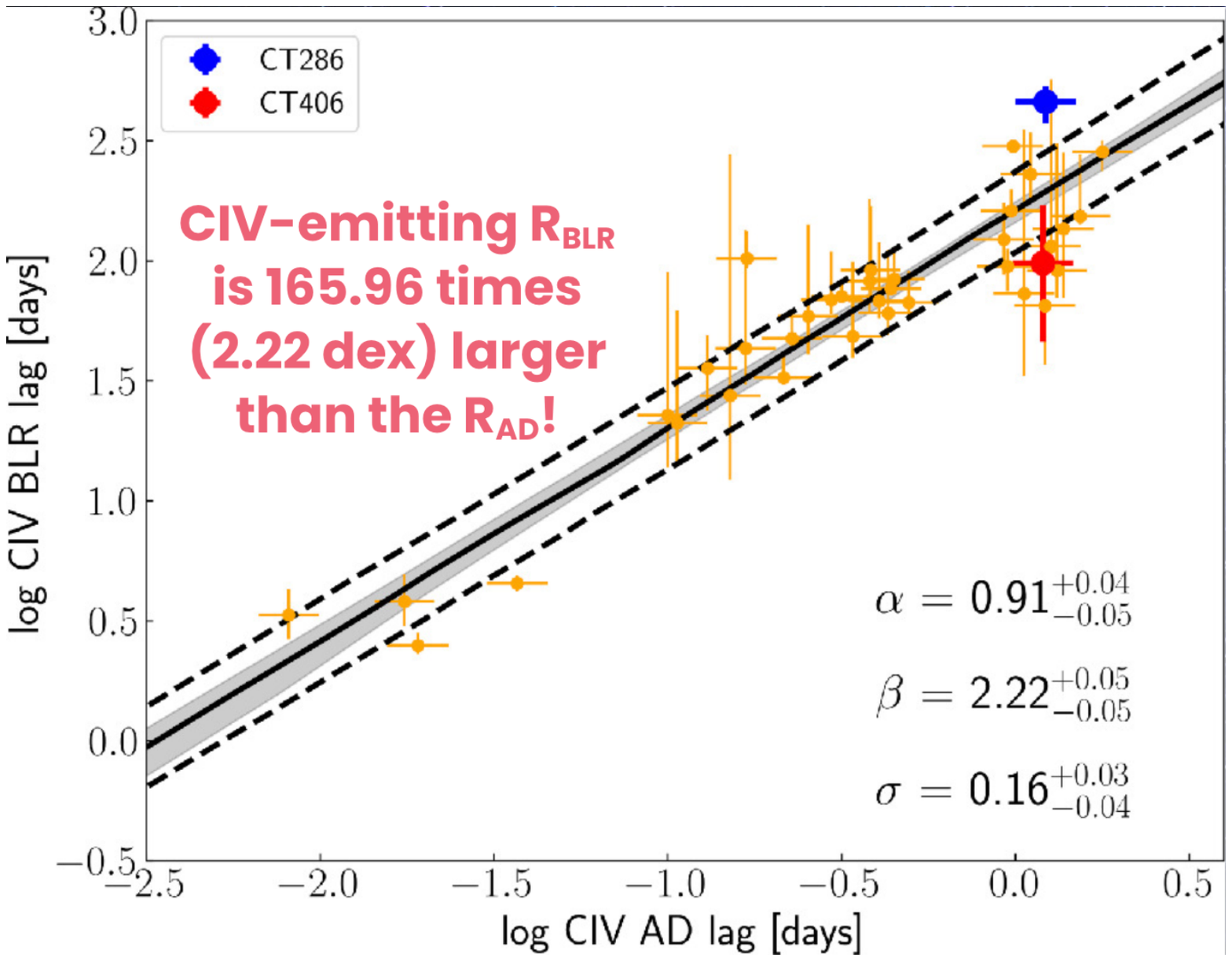}
    \caption{Recovered $R_{\mathrm{BLR}}$--$R_{\mathrm{AD}}$ relation; Courtesy: \citet{PandaFrancisco2024ApJL}. The solid black line represents the mean of the posterior probability distributions, while the shaded region denotes the corresponding $1\sigma$ confidence interval. The best-fit slope ($\alpha$), intercept ($\beta$), and intrinsic scatter ($\sigma$)  are reported together with their $1\sigma$ uncertainties. The dashed lines indicate the mean model predictions at the upper and lower bounds when the intrinsic scatter is incorporated. The positions of the two studied sources, CT286 and CT406, which belong to the class of high-accreting sources, are marked by the blue and red circles, respectively.}
    \label{fig:ad-blr}
\end{figure}

To address this, we utilized the photometric reverberation mapping (PRM) technique as a rapid, cost-effective alternative to spectroscopic RM for measuring accretion-disk (AD) sizes in high-redshift quasars via the C {\sc iv} emission line \citep{PandaFrancisco2024ApJL}. Using carefully chosen medium-band filters that isolate line-free continuum regions (with $\leq$2\% BLR contamination), a high-cadence monitoring campaign at the meter-class telescopes recovered rest-frame AD time delays with 10–15\% precision. This method proved $\sim$166 times faster than traditional BLR-based RM (Figure \ref{fig:ad-blr}).\\

Simulations based on thin-disk reprocessing ($\tau \propto \lambda^{4/3}$) coupled with high signal-to-noise ratios ($\sim$100) observations, demonstrate that modest-aperture facilities can deliver robust PRM results. The analysis yielded a radius–luminosity (R$_{\rm AD}$–L$_{\rm 1350}$) relation with $\beta \approx 0.5$, consistent with photoionization theory. By calibrating a scaling relation between AD and BLR sizes, black-hole masses can be estimated with $\sim$23\% uncertainty - an efficient alternative where direct BLR reverberation is impractical.\\

Building on this, \citet{FPN2025A&AL} demonstrated the technique during a six-month campaign of QSO J0455–4216 ($z = 2.662$), mapping an accretion disk beyond the cosmic noon for the first time. The recovered delay spectrum followed the expectations of a Shakura–Sunyaev thin disk \citep{1973A&A....24..337S} irradiated in a lamp-post geometry, yielding a mean emissivity radius of $4.75^{+1.12}_{-1.05}$ light-days in the observer’s frame (1.29 light-days rest frame) and implying a black hole mass of $\sim$9 $\times 10^8$ M$\odot$.\\

In summary, traditional C {\sc iv}-based BLR lags at $z \sim 2$–3 required nearly two decades of monitoring and revealed BLR sizes corresponding to $\sim$16-year lags. In contrast, PRM can constrain AD sizes with only 4–5 months of dense monitoring, scale these to BLR sizes, and, with the aid of single-epoch spectroscopy and the virial relation, estimate black hole masses and Eddington ratios. This approach transforms the feasibility of high-redshift RM: small ground-based telescopes equipped with medium- and narrow-band filters can now probe accretion disks and black hole growth for hundreds to thousands of quasars across cosmic time, opening a rewarding avenue for future studies.

\section{Why we should care about accretion-dependent SEDs and closing remarks}

BHAR is straightforward to estimate; however, it depends crucially on the determination of the net bolometric output, or bolometric luminosity (L$_{\rm bol}$). The estimation of the L$_{\rm bol}$ is not trivial and requires estimating the area under the broad-band spectral energy distribution (SED), although a crude estimate can be derived assuming a constant scaling \citep{Richards_2006ApJS..166..470R} or a parametric `bolometric correction' \citep{Netzer_2019MNRAS.488.5185N} based on mean SEDs derived from composite AGN spectra. However, the overabundance of overly massive black holes at high-z \citep{massiveBH_2024ApJ...966L..30M, massiveBH_2025MNRAS.537.2323H} and them suggested to accrete at super-Eddington limits \citep{PandaMarziani2023FrASS, 2025Univ...11...69M} is closely tied to our reliance on scaling relations and mean SEDs built from local samples, the latter being limited and originally made from the general population of quasars. Therefore, not only do we need accretion-dependent SEDs, but to infer bulk statistical properties of these highly-accreting AGNs and their neighborhood, we are in dire need to update our AGN SED databases, especially at higher redshifts - as we push forth to spatially resolve BLR in quasars at z=4 \citep{2025arXiv250913911G}.\\

\begin{figure}
    \centering
    \includegraphics[width=\linewidth]{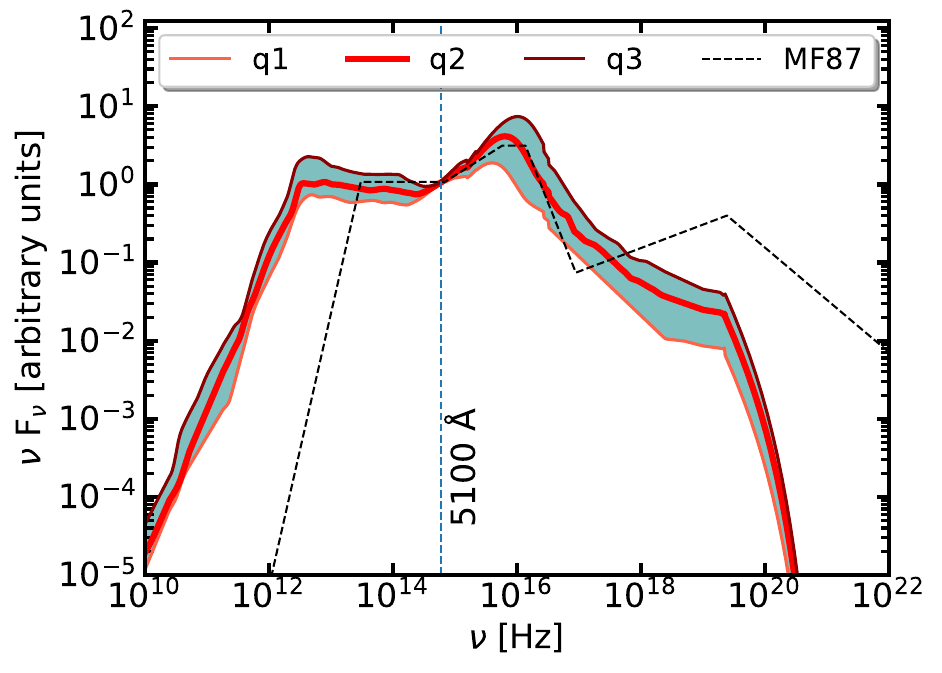}
    \caption{Spectral Energy Distributions (SEDs) from \citet{Garnica_2025MNRAS.540.3289G} for high-Eddington accreting (xA) quasars. The median SED (q2) and the inter-quartile range defined by the lower (q1) and higher (q3) SEDs are also highlighted. For reference, the often used AGN SED from \citet{MF_1987ApJ...323..456M} is also shown, demonstrating the variations in the broad-band SED shape relative to the Fe {\sc ii} strength-based SEDs.}
    \label{fig:seds}
\end{figure}

To this end, \citet{Garnica_2025MNRAS.540.3289G} presented semi‑empirical spectral energy distributions (SEDs) tailored to extreme, high‑accretion (xA) quasars. By integrating extensive optical survey data with multi‑frequency archival data, we constructed median SEDs that capitalize on the pronounced spectral homogeneity observed among super‑Eddington candidates \citep{PandaMarziani2023FrASS}. The SED construction is performed in three spectral regimes - radio to NIR, optical-UV dominated by the accretion‑disk, and in X‑ray - each normalized at 5100\AA\ to facilitate direct comparison with conventional templates. A systematic juxtaposition with the canonical Mathews \& Ferland (MF87) SED underscores substantive divergences, particularly in the high‑energy domain (see Figure \ref{fig:seds}). To evaluate the ramifications for broad‑line region (BLR) diagnostics and how these SEDs affect the emitting regions, we employed {\tt cloudy} photoionization simulations, demonstrating that the newly derived SED reproduces the extreme BLR conditions characteristic of xA quasars: markedly low ionization parameters, elevated gas densities, and supersolar metallicities. These physical conditions naturally account for the observed suppression of C {\sc iv} relative to H$\beta$ and the pronounced Fe {\sc ii} emission. In summary, the work delivers a data‑driven, semi‑empirical xA SED that diverges from traditional quasar templates and validates its influence on BLR physical‑parameter determinations.\\

We provide {\tt cloudy}-ready template median SED (q2) for the RQ extreme population A (drawn from the systematic analysis of 139 R$_{\rm Fe}$-rich sources), along with the first and third quartiles (q1 and q3) for the community. \citep{Garnica_2025MNRAS.540.3289G} suggests that for these xA SEDs characterized by high-to-super accretion mode, the bolometric output is more pronounced relative to the mean quasar SED from \citet{Richards_2006ApJS..166..470R}, between 1.2 - 1.7 times higher\footnote{the bolometric corrections for the q1, q2 and q3 SEDs (luminosity based) are 11.16, 15.70, and 11.00, respectively.}. \\

In summary, we see that the Eddington ratio has come to our rescue in more than one way, and with the growing interest over the last decades and coming years promising more advancements and challenges, we are well-poised for many more interesting AGN-related discoveries.

\acknowledgements
S.P. is supported by the international Gemini Observatory, a program of NSF NOIRLab, which is managed by the Association of Universities for Research in Astronomy (AURA) under a cooperative agreement with the U.S. National Science Foundation, on behalf of the Gemini partnership of Argentina, Brazil, Canada, Chile, the Republic of Korea, and the United States of America. SP acknowledges the entire organizing committee of the 15$^{\rm th}$ Serbian Conference on Spectral Line Shapes in Astrophysics, for their invitation to this memorable conference. S.P. acknowledges Michael Eracleous and Ed Cackett for their helpful contributions and feedback. M.S. acknowledges support from the European Research Council (ERC) under the European Union’s Horizon 2020 research and innovation program (grant agreement number 950533), and the Israel Science Foundation (grant number 1849/19). F.P.N. gratefully acknowledges the generous and invaluable support of the Klaus Tschira Foundation and funding from the European Research Council (ERC) under the European Union's Horizon 2020 research and innovation program (grant agreement No 951549).
\bibliography{demo_caosp310}

\clearpage

\end{document}